\documentclass[name,preprint,review,12pt]{elsarticle}
\usepackage{amssymb}  % mathematical symbols
\usepackage{amsmath}  % equation environments
\usepackage{CJKutf8}  % chinese
\usepackage{lineno}   % line number for review, linenumbers/pagewiselinenumbers
\usepackage{bm} % 导入 bm 宏包
\usepackage{float}
\usepackage{tabularx}  
\usepackage{ragged2e}  % 用于表格内自动换行的对齐
\usepackage{booktabs}  % 用于画漂亮的三线表
\usepackage{multirow}
\usepackage{caption}
\usepackage{subcaption}
\usepackage{graphicx}   % 插入图片
\usepackage{hyperref}

\makeatletter
\def\ps@pprintTitle{
  \let\@oddhead\@empty
  \let\@evenhead\@empty
  \let\@oddfoot\@empty
  \let\@evenfoot\@empty
}
\makeatother

\begin{document}
\begin{CJK*}{UTF8}{gbsn}  % chinese

% Title + Abstract + Keywords -----------------------------------------------
\begin{frontmatter}
  \title{Model-Driven Conditional Fourier Neural Operator for Spectrum-Consistent Synthetic Turbulence Generation}

  \author{Hongyuan Lin}
  \author{Shizhao Wang\corref{cor}}
  \ead{wangsz@lnm.imech.ac.cn}
  \cortext[cor]{Corresponding author.}
  \affiliation{organization={Institute of Mechanics, Chinese Academy of Sciences},
              % addressline={},
              % city={Beijing},
              postcode={100190},
              state={Beijing},
              country={China}}

  \begin{abstract}
    This short note proposes a model-driven conditional Fourier neural operator (MD-CFNO) for synthetic turbulence generation. Spectrum-consistent synthetic turbulence is essential for inflow boundary construction in computational fluid dynamics and for broadband aeroacoustic noise prediction. Data-driven turbulence synthesis with neural networks has emerged as a promising direction. However, generating flow fields that match prescribed energy spectra across wide physical regimes remains challenging. Existing data-driven methods typically rely on expensive reliable datasets with limited generalization and are prone to regression-to-the-mean when trained in the spatial domain.
    To address these issues, the MD-CFNO is proposed with three components: a model-driven data construction strategy is adopted to improve interpretability and broaden the generalizable parameter regime; conditional stochastic generation is integrated into the Fourier neural operator architecture to alleviate regression-to-the-mean effects; and a composite loss is introduced to accelerate convergence and enhance spectral fidelity.
    Results show that the proposed MD-CFNO generates spectrum-consistent synthetic turbulence and achieves robust performance under both interpolation and out-of-distribution extrapolation conditions.
    This study provides a model-driven perspective on synthetic turbulence, showing the advantages of Fourier neural operators for conditional generation.
  \end{abstract}

  % \begin{graphicalabstract}
  % %\includegraphics{grabs}
  % \end{graphicalabstract}
 
% %   % Elsevier
%     \begin{highlights}
%     \item The proposed MD-CFNO realizes controllable, spectrum-consistent synthetic turbulence generation.
%     \item Model-driven strategy internalizes theoretical spectral laws without expensive simulation costs.
%     \item Wavenumber-domain integration eliminates the regression-to-the-mean in stochastic generation.
%     \item Superior extrapolation capability is achieved under out-of-distribution conditions.
%     \end{highlights}

\end{frontmatter}

% Main Body -----------------------------------------------------------------
\section{Introduction}
\label{sec:1-Intro}  
    % Importance of Present Research
    % Motivation: why synthetic turbulence?
    The objective of this work is to propose a model-driven conditional Fourier neural operator (MD-CFNO) for spectrum-consistent synthetic turbulence generation. Generative modeling is a central paradigm in modern artificial intelligence, aiming to approximate the latent probability distribution of complex data and to sample from it to produce new realizations consistent with the underlying distribution \cite{du2024conditional, wu2020enforcing}. Substantial progress has been made in high-resolution image synthesis \cite{saharia2022photorealistic}, protein structure prediction \cite{wu2024protein}, and molecular design \cite{huang2024dual}. Extending these models to physical problems remains nontrivial, in particular when essential structural properties and physical attributes must be preserved during generation \cite{wang2025fundiff}.
    
    % Previous and/or Current Research and Contributions
    Recent efforts have extended modern generative modeling to synthetic turbulence generation, with representative data-driven frameworks including variational autoencoders (VAEs) \cite{fukami2019synthetic, yousif2022physics}, generative adversarial networks (GANs) \cite{kim2021unsupervised, kim2020deep}, and diffusion models \cite{du2024conditional, dong2025data, shu2023physics, li2024synthetic}. First, autoencoder-based methods were adopted for latent representation learning and flow synthesis: Fukami et al.~\cite{fukami2019synthetic} trained a hybrid autoencoder on fully developed channel flow at a friction Reynolds number of $Re_{\tau}=180$ to construct a machine-learning inflow generator, while Yousif et al.~\cite{yousif2022physics} combined multiscale convolutional autoencoders with long short-term memory networks to reproduce spatiotemporal statistics at $Re_{\tau}=180$ and $550$. Second, GAN-based models improved textural realism and enhanced Reynolds-number generalization. Kim et al.~\cite{kim2020deep} coupled GANs with recurrent neural networks to generate turbulent fields at unseen Reynolds numbers after training on $Re_{\tau}=180, 360,$ and $540$. Third, diffusion models have recently emerged as a competitive paradigm for turbulence synthesis \cite{yang2023denoising, shu2023physics, gao2024bayesian, li2024synthetic, liu2025confild}. In particular, Li et al.~\cite{li2024synthetic} developed a diffusion-based framework for synthetic Lagrangian turbulence that reproduces Lagrangian statistics from large scales down to the inertial-viscous range, captures local scaling behavior near the Kolmogorov time scale, and improves small-scale fidelity relative to Wasserstein GAN baselines. The diffusion-based CoNFiLD-inlet model \cite{liu2025confild} parameterized inflow conditions and showed generalization over $Re_{\tau}=10^3$-$10^4$.

    % Research Gaps/Problem/Question/Challenges
    Despite these advancements, achieving conditionally controllable turbulence synthesis faces two primary challenges. First, traditional neural networks trained in the spatial domain are prone to regression-to-the-mean \cite{fukami2019synthetic}, resulting in overly smoothed flow fields that fail to effectively capture detailed turbulent structures, leading to deviations from the theoretical energy spectrum. Second, data-driven turbulence generators are often constrained by the high cost of direct numerical simulation (DNS) or sparse experimental measurements \cite{kim2020deep,liu2025confild}, resulting in training datasets that cover only a limited portion of the parameter space and consequently impairing generalization.

    % Methodological Gap    
    % Objectives and Significance
    % Contributions:  
    To address these issues, we propose the MD-CFNO for synthetic turbulence generation. From an operator-learning perspective, the task is posed as a nonlinear mapping from Gaussian white noise to a velocity field that satisfies the prescribed turbulence energy spectrum, with turbulent kinetic energy and dissipation rate used as explicit control variables for spectrum-consistent conditioning. The main contributions are summarized as follows:
    \begin{enumerate}
      \item[(1)] A model-driven strategy based on the random Fourier model is introduced to replace high-cost DNS data, embedding spectral priors into the data construction stage to enhance interpretability and extend the generalizable parameter regime.
      \item[(2)] A conditional stochastic generation method is combined with the Fourier neural operator as a conditional stochastic generator for synthetic turbulence generation. The global integral operator in the wavenumber domain is used to mitigate the regression-to-the-mean that often arises in convolutional architectures for random-field generation.
      \item[(3)] A multi-constraint loss is designed by incorporating explicit wavenumber-domain and spectrum consistency constraints, which improves training convergence and helps enforce adherence to the prescribed energy spectrum.
    \end{enumerate}

%=================================================================================================================
\section{Methodology}
\label{sec:2-Method}   
    
    %=================================================================================================================
    \subsection{Overview}
    \label{sec:2.1-Overview}
    This work aims to develop an operator-learning-based generative model that rapidly synthesizes turbulent velocity fields $\bm{u}(\bm{x})$ with a prescribed energy spectrum, conditioned on the turbulent kinetic energy ($\mathit{TKE}$) and the dissipation rate ($\varepsilon$). The turbulence generation problem is formulated as a nonlinear conditional mapping from a Gaussian white-noise space to a physical turbulent velocity field. The corresponding generative operator $\mathcal{G}_\theta$ is defined as
    \begin{equation}
        \mathcal{G}_\theta : \bigl(z(\bm{x}),\,\bm{c}\bigr) \;\mapsto\; \bm{u}_{\mathrm{pred}}(\bm{x}),
    \label{eq:operator_mapping}
    \end{equation}
    where $z(\bm{x})$ is a random Gaussian white-noise field accounting for turbulence stochasticity; $\bm{c}$ is the physical condition vector, including $\mathit{TKE}$, $\varepsilon$, and the spectrum-related coefficients $\alpha$ and $\kappa_e$ derived from them; $\bm{u}_{\mathrm{pred}}(\bm{x})$ is the synthesized velocity field; and $\theta$ denotes the trainable parameters of the neural model.

\begin{figure}[!htbp]
      \centering
      \includegraphics[width=\linewidth]{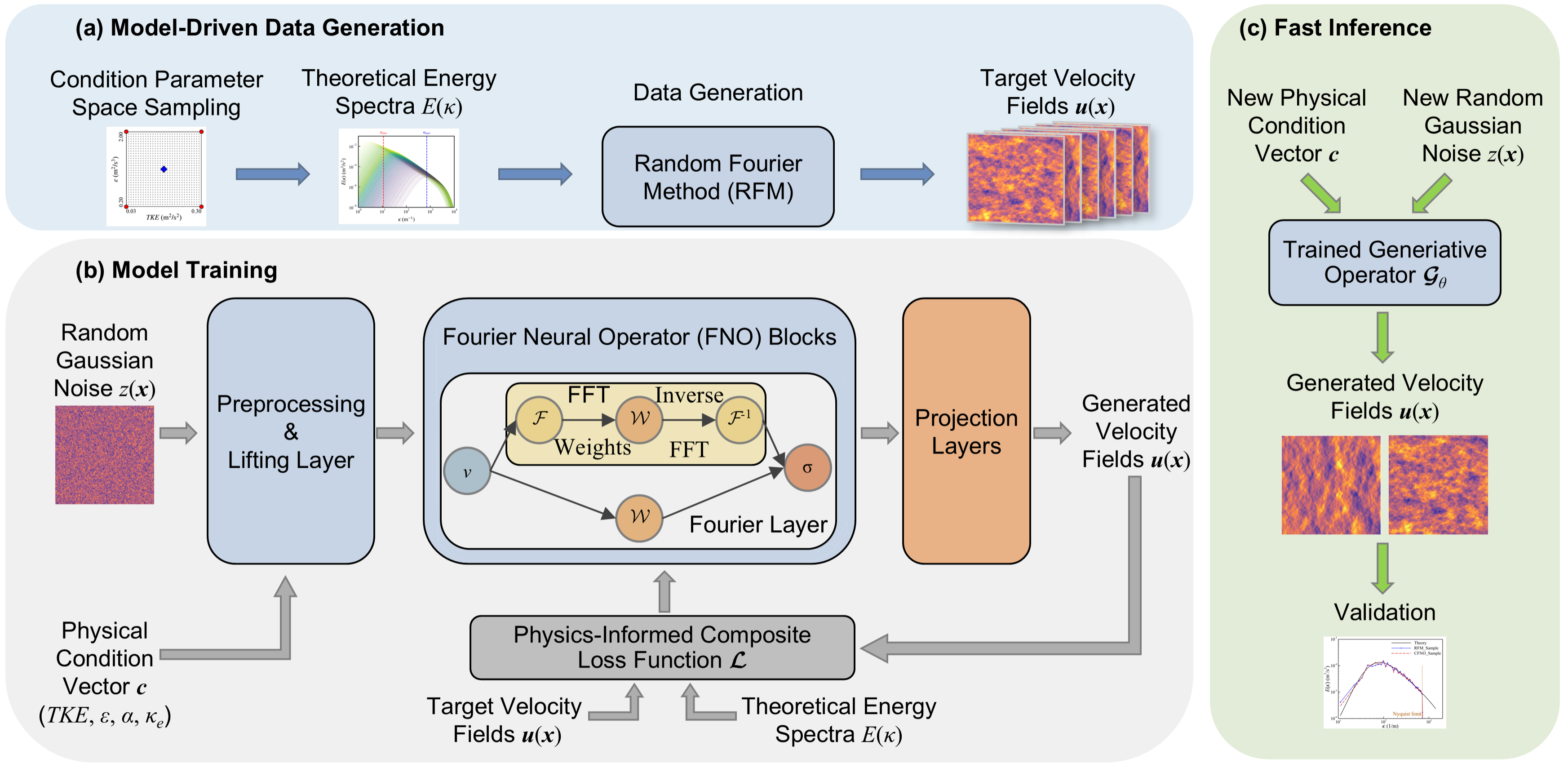}
      \caption{Schematic overview of the proposed MD-CFNO framework for spectrum-consistent synthetic turbulence generation.}
      \label{fig:flowchart}
    \end{figure}

    An overview of the proposed approach is shown in Fig.~\ref{fig:flowchart}. First, a model-driven data construction strategy is adopted by replacing expensive DNS with the random Fourier model (RFM) to build a model-embedded dataset that covers a continuous parameter space, thereby establishing a theoretical mapping between the input condition $\bm{c}$ and the target field $\bm{u}_{\mathrm{target}}$. Second, based on a conditional Fourier neural operator architecture, the physical condition $\bm{c}$ is embedded into the neural model to alleviate the locality limitation and regression-to-the-mean behavior of convolutional networks. Finally, a multi-constraint loss is designed by combining spatial-domain reconstruction, wavenumber-domain constraints, and statistical spectral matching, which improves training convergence.
    
    %=================================================================================================================
    \subsection{Model-Driven Strategy}
    \label{sec:2.2-Model-Driven}
    A model-driven training strategy is proposed to allow the neural operator to internalize spectral characteristics across different physical conditions while avoiding the prohibitive cost of direct numerical simulation (DNS).
    Specifically, the random Fourier model (RFM) is used to construct a training dataset that satisfies prescribed physical constraints. In this approach, the target velocity field is synthesized by superposing Fourier modes in the wavenumber domain according to a specified energy spectrum:
    \begin{equation}
        \bm{u}_{\mathrm{target}}(\bm{x}) = 2 \sum_{m=1}^{M} u_m \cos(\bm{\kappa}_m \cdot \bm{x} + \psi_m) \bm{\sigma}_m,
        \label{eq:velocity_field}
    \end{equation}
    where $\bm{\kappa}_m$ is the wavenumber vector, $\psi_m$ is a random phase, and $\bm{\sigma}_m$ is a direction vector chosen to enforce the divergence-free condition. The amplitude $u_m$ is determined by the target energy spectrum $E(\kappa)$. The model-driven strategy is illustrated using the von K\'arm\'an--Pao spectrum \cite{Bailly1999, Karman1948, Pao1965}:
    \begin{equation}
        E(\kappa) = \alpha \frac{2\mathit{TKE}}{3\kappa_e} \left( \frac{\kappa}{\kappa_e} \right)^4 \left[ 1 + \left( \frac{\kappa}{\kappa_e} \right)^2 \right]^{-17/6} \exp \left[ -2 \left( \frac{\kappa}{\kappa_\eta} \right)^2 \right],
        \label{eq:karman_spectrum}
    \end{equation}
    where $\alpha$ is an energy scaling factor and $\kappa_e$ is the characteristic wavenumber of the energy-containing range that controls the spectral peak; $(\alpha, \kappa_e)$ are derived from the input condition $(\mathit{TKE}, \varepsilon)$ \cite{Guglielmi2025}. The Kolmogorov wavenumber is given by $\kappa_\eta = (\varepsilon/\nu^3)^{1/4}$. By continuously varying $(\mathit{TKE}, \varepsilon)$ and the associated derived parameters, Eq.~\ref{eq:karman_spectrum} provides direct control of the energy spectrum of the synthesized turbulence. This strategy generates turbulence fields covering the prescribed range of conditions and ensures spectrum consistency of the training data at each parameter setting.

   %=================================================================================================================
    \subsection{Conditional Fourier Neural Operator}
    \label{sec:2.3-MD-CFNO}    
    In order to enable conditional generation under prescribed condition $\bm{c}$, a conditional Fourier neural operator is developed based on the model-driven strategy described above. Conditional neural networks have been widely used for label-guided generation in image synthesis \cite{saharia2022photorealistic} and for conditional spatiotemporal field reconstruction \cite{du2024conditional}, whereas conditional generation of random fields within an operator-learning setting remains relatively less explored. Here, the conditional generation is combined with Fourier neural operator. Its global integral operator in the wavenumber domain facilitates learning of multiscale spectral characteristics of turbulence and thus alleviate the limitations of pointwise approximation in the spatial domain \cite{li2020fourier, wang2025fundiff}.

    Given the Gaussian white-noise field $z(\bm{x})$ and the condition vector $\bm{c}$ defined in Section~\ref{sec:2.1-Overview}, the spatially augmented input is constructed as $\bm{a}(\bm{x})=[z(\bm{x});\mathcal{B}(\bm{c})]$, where $\mathcal{B}(\cdot)$ broadcasts $\bm{c}$ to match the spatial grid. The conditional operator $\mathcal{G}_\theta$ in eq.\ref{eq:operator_mapping} is parameterized by a conditional Fourier neural operator with a lifting layer $P$, $L$ stacked Fourier integral layers $\{\mathcal{L}_l\}_{l=0}^{L-1}$, and a projection layer $Q$. The forward pass of the proposed architecture is illustrated in Fig.~\ref{fig:flowchart} and summarized as
    \begin{equation}
    v_0 = P(\bm{a(\bm{x})}), 
    \qquad 
    v_L = \bigl(\mathcal{L}_{L-1}\circ\cdots\circ\mathcal{L}_0\bigr)(v_0),
    \qquad
    \bm{u}_{\mathrm{pred}} = u_{\mathrm{rms}}\, Q(v_L),
    \label{eq:mdcfno_overall}
    \end{equation}
    where the stacked composition explicitly highlights the iterative refinement mechanism enabled by the Fourier integral layers, and the amplitude is calibrated by $u_{\mathrm{rms}}=\sqrt{2\mathit{TKE}/3}$.
    
    Each Fourier integral layer $\mathcal{L}_l$ updates the latent feature field via a residual-type spectral mixing operator,
    \begin{equation}
    v_{l+1}(\bm{x})
    =
    \sigma\!\left(
    W_l v_l(\bm{x})
    +
    \mathcal{F}^{-1}\!\left(
    R_l(\bm{\kappa})\odot \mathcal{F}(v_l)(\bm{\kappa})
    \right)(\bm{x})
    \right),
    \qquad l=0,\dots,L-1,
    \label{eq:fno_layer}
    \end{equation}
    where $\sigma$ is the Gaussian error linear unit (GELU) activation, $W_l$ denotes a learnable linear operator in the physical space, and $\mathcal{F}$ and $\mathcal{F}^{-1}$ are the fast Fourier transform and its inverse. The wavenumber-domain kernel $R_l$ is complex-valued and defined on truncated modes $|\bm{\kappa}|\le\bm{\kappa}_{\max}$, providing global interactions across resolved Fourier modes.
    
    %=================================================================================================================
    \subsection{Physics Constrained Loss Function}
    \label{sec:2.4-Loss}
    A composite loss function $\mathcal{L}$ is designed to guide learning toward the correct sample distribution while reducing the statistical smoothing effect induced by pointwise training objectives, which is constructed with three components:
    \begin{equation}
        \mathcal{L} = \lambda_{\mathrm{spat}} \mathcal{L}_{\mathrm{spat}} + \lambda_{\mathrm{wave}} \mathcal{L}_{\mathrm{wave}} + \lambda_{\mathrm{spec}} \mathcal{L}_{\mathrm{spec}},
    \end{equation}
    where $\lambda_{\mathrm{spat}}$, $\lambda_{\mathrm{wave}}$, and $\lambda_{\mathrm{spec}}$ are the weights for spatial-domain reconstruction, wavenumber-domain amplitude matching, and energy spectrum consistency, respectively. The loss terms are defined as follows.
    
    (1) The spatial-domain reconstruction loss $\mathcal{L}_{\mathrm{spat}}$ supervises the large-scale structure of the velocity field. A standard mean-squared error is used to enforce agreement between the predicted field $\bm{u}_{\mathrm{pred}}$ and the target field $\bm{u}_{\mathrm{target}}$:
    \begin{equation}
        \mathcal{L}_{\mathrm{spat}} = \frac{1}{B} \sum_{i=1}^{B} \| \bm{u}_{\mathrm{pred}}^{(i)} - \bm{u}_{\mathrm{target}}^{(i)} \|_2^2,
    \end{equation}
    where $B$ is the batch size configured for the training process and $\|\cdot\|_2^2$ denotes the squared $L_2$ norm over the full field.
    
    (2) The wavenumber-domain amplitude loss $\mathcal{L}_{\mathrm{wave}}$ explicitly constrains Fourier amplitudes to mitigate the insensitivity of spatial-domain MSE to high-wavenumber errors. This term computes the mean-squared error between the amplitude spectra of the predicted and target fields in the wavenumber domain:
    \begin{equation}
        \mathcal{L}_{\mathrm{wave}} = \frac{1}{B \cdot C} \sum_{i=1}^{B} \sum_{c=1}^{C} \| |\mathcal{F}(u_{pred, c}^{(i)})| - |\mathcal{F}(u_{target, c}^{(i)})| \|_2^2,
    \end{equation}
    where $\mathcal{F}(\cdot)$ denotes the fast Fourier transform, $|\cdot|$ denotes the complex modulus, and $C$ is the number of velocity components.
    
    (3) The energy spectrum consistency loss $\mathcal{L}_{\mathrm{spec}}$ is the core statistical constraint that enforces agreement with the prescribed spectrum. Rather than pointwise matching, $\mathcal{L}_{\mathrm{spec}}$ constrains the isotropically averaged energy spectrum of the predicted field, $E_{pred}(\kappa)$, to match the target von K\'arm\'an--Pao spectrum $E_{target}(\kappa)$. Since the spectrum decays rapidly with increasing $\kappa$, a wavenumber-weighted logarithmic $L_1$ loss is adopted:
    \begin{equation}
        \mathcal{L}_{\mathrm{spec}} = \frac{1}{B \cdot K} \sum_{i=1}^{B} \sum_{k=1}^{K} w_k \left| \log \left( E_{pred}^{(i)}(\kappa_k) + \epsilon \right) - \log \left( E_{target}^{(i)}(\kappa_k) + \epsilon \right) \right|,
        \label{eq:loss_spec}
    \end{equation}
    where $K$ is the number of effective wavenumber bins determined by the spatial resolution, $\epsilon$ is a small constant for numerical stability, and the weight is defined as $w_k = \kappa_k / \kappa_{min}$ to increase emphasis toward higher wavenumbers and compensate for the decay of spectral magnitude.

%=================================================================================================================
\section{Dataset Configuration and Implementation}
\label{sec:3-Data}     

    %=================================================================================================================
    \subsection{Dataset Configuration}
    In order to assess the generalization capability of the proposed MD-CFNO across different physical conditions, a synthetic turbulence dataset is constructed using the model-driven data strategy described in Sec.~\ref{sec:2.2-Model-Driven}. The physical parameter space is spanned by the turbulent kinetic energy ($\mathit{TKE}$) and the dissipation rate ($\varepsilon$). Once a condition $(\mathit{TKE}, \varepsilon)$ is specified, the spectral shape parameters $\alpha$ and $\kappa_e$ in Eq.~\ref{eq:karman_spectrum} are uniquely determined through prescribed physical relations \cite{Guglielmi2025}, establishing a one-to-one mapping between the physical condition and the corresponding target spectrum.

\begin{figure}[!htb]
        \centering
        \begin{subfigure}[b]{0.48\linewidth}
            \centering
            \includegraphics[width=\linewidth]{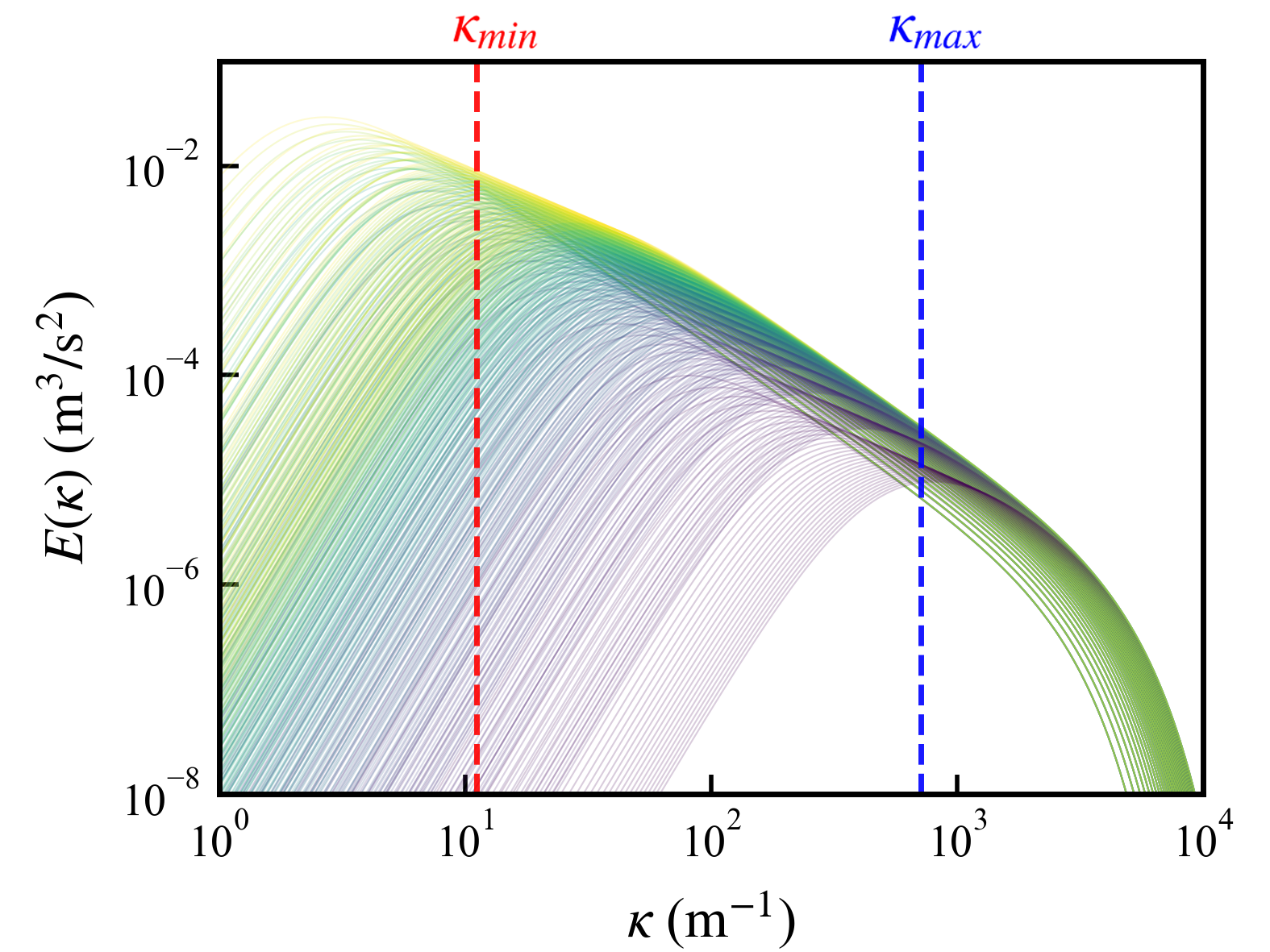}
            \caption{Evolution of energy spectrum.}
        \end{subfigure}
        \hfill 
        \begin{subfigure}[b]{0.48\linewidth}
            \centering
            \includegraphics[width=\linewidth]{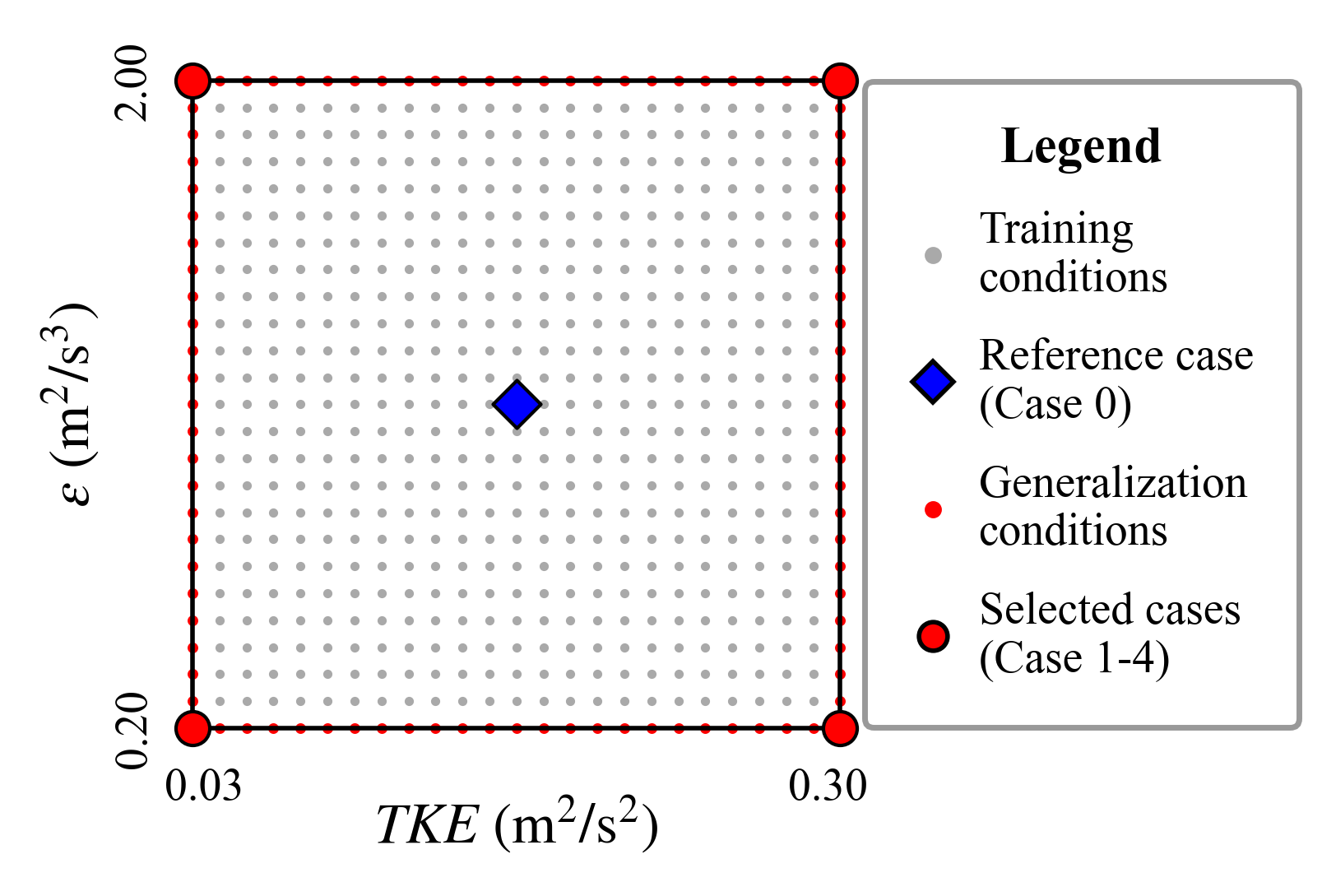}
            \caption{Parameter space and data split.}
        \end{subfigure}
        \caption{Visualization of the model-driven data strategy and parameter space coverage.}
        \label{fig:SpectrumSampling}
    \end{figure}

    Following our previous study \cite{lin2025}, the domain size is set to $L = 0.18\pi$~m with a grid resolution of $128 \times 128$. The parameter ranges are selected such that, for all combinations, the key spectral features of the associated energy spectrum, from the energy-containing range to the dissipation range, lie within the effective wavenumber bound supported by the domain size and the Nyquist limit. Accordingly, the parameter space is sampled linearly as $\mathit{TKE} \in [0.03, 0.30]~\mathrm{m}^2/\mathrm{s}^2$ and $\varepsilon \in [0.20, 2.00]~\mathrm{m}^2/\mathrm{s}^3$. The resulting integral-scale Reynolds number, $Re_L = \mathit{TKE}^2/(\nu \varepsilon)$, spans from 45 to 45,000. This specific regime is targeted to capture the most distinct spectral evolution features resolvable within the fixed grid, rather than representing the intrinsic capability limit of the proposed method. The corresponding evolution of the target spectra over the parameter space is shown in Fig.~\ref{fig:SpectrumSampling}(a), where dashed lines indicate the effective wavenumber bounds imposed by the domain size and the Nyquist limit \cite{lin2025}. Within the selected ranges, the spectra exhibit substantial variability over the effective bound: the peak wavenumber $\kappa_e$, associated with the largest energy-containing scale, shifts over $[1.67, 528.09]~\mathrm{m}^{-1}$, and the peak magnitude varies across approximately $10^{-4}$ to $10^{-2}$. These variations provide a stringent test for learning and generating multiscale spectral distributions under finite-grid constraints.
    
    To obtain a discrete set of physical conditions, the continuous parameter space is discretized into $32 \times 32 = 1{,}024$ condition points, as illustrated in Fig.~\ref{fig:SpectrumSampling}(b). For each condition point, $N_{\mathrm{seed}}=4$ statistically independent realizations are generated, yielding a total of 4,096 flow-field samples. This sampling strategy, combining multiple conditions with multiple realizations, balances parameter-space coverage, statistical consistency, and stochastic variability, and facilitates learning of turbulence statistics that are not tied to specific realizations.

    %=================================================================================================================
    \subsection{Implementation Details}
    All models are implemented in PyTorch and trained on a single NVIDIA GeForce RTX 3090 GPU. The physical condition vector $\bm{c}$ is normalized to $[0,1]$ using Min--Max scaling to improve optimization stability, where normalization statistics are computed from the training set only and then applied consistently to validation and test data. The data split is performed at the level of physical condition points. The parameter space contains $32\times 32=1024$ condition points, each with $N_{\mathrm{seed}}=4$ realizations. To assess out-of-distribution generalization in parameter space, all condition points on the four edges of the parameter grid (a total of 124 points) are reserved as an out-of-distribution test set and excluded from training, as illustrated in Fig.~\ref{fig:SpectrumSampling}(b). The remaining interior condition points form the training pool, from which 10\% of the condition points are randomly selected as a validation set for monitoring training and selecting hyperparameters. Optimization is performed using Adam with an initial learning rate of $1\times 10^{-3}$ and a cosine-annealing schedule with $T_{\max}=30$. Training is conducted for 30 epochs with a batch size of 32. The loss weights are set to $\lambda_{\mathrm{spat}}=\lambda_{\mathrm{wave}}=1.0$, and the spectrum consistency term is emphasized by setting $\lambda_{\mathrm{spec}}=10.0$. The total training time is approximately 2 hours.

%=================================================================================================================
\section{Results and Discussion}
\label{sec:4-Results}    
    
    %=================================================================================================================
    \subsection{Performance Validation}
    \label{subsec:Comparison}
    
    This section first validates the architectural advantage of the proposed MD-CFNO through a comparison with a representative convolutional baseline, and then evaluates generation quality and computational efficiency against the traditional RFM.
    
\begin{figure}[!htbp]
      \centering
      \includegraphics[width=0.6\linewidth]{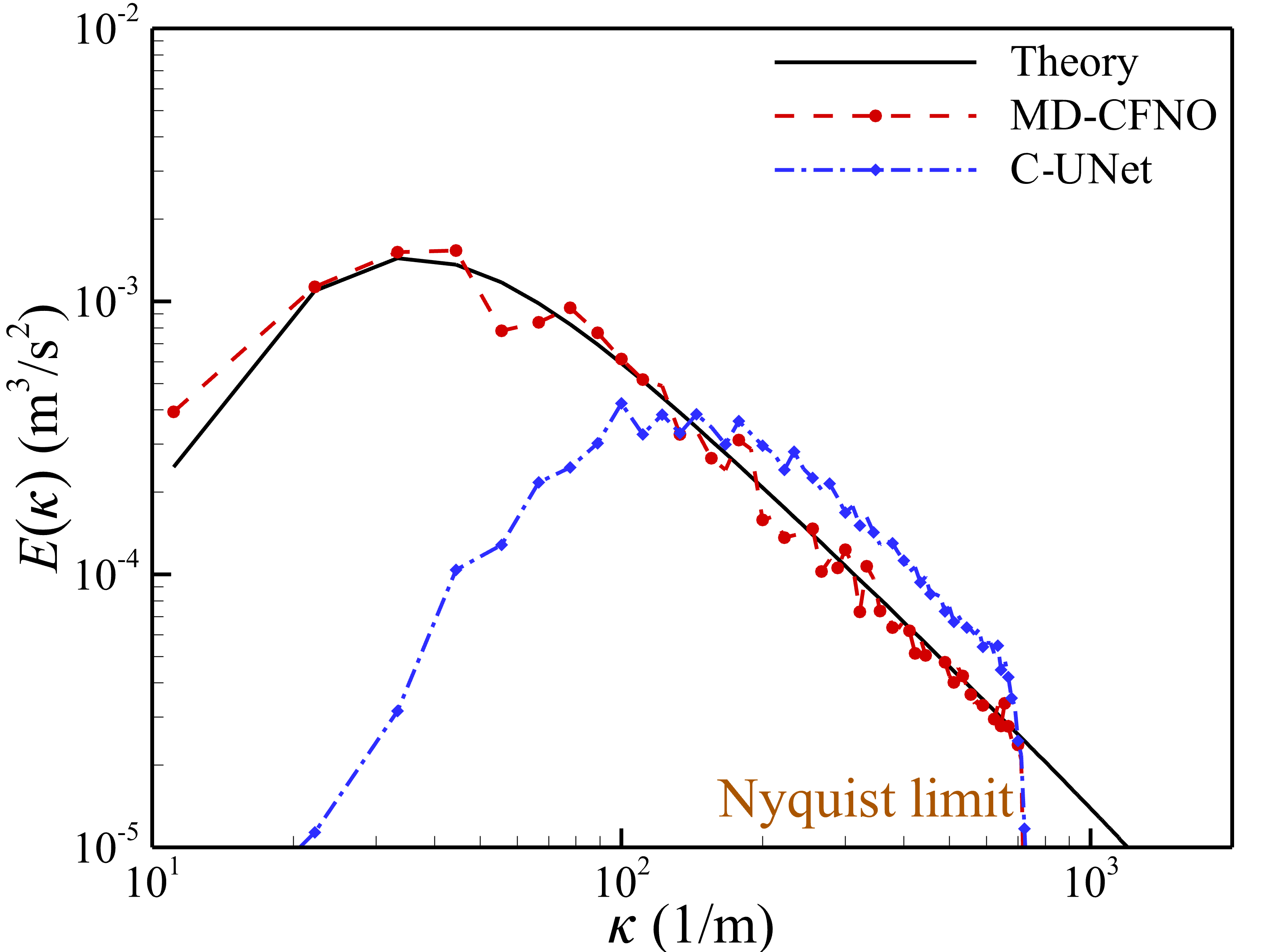}
      \caption{Spectral accuracy comparison between the proposed MD-CFNO and C-UNet at the reference condition (Case 0).}
      \label{fig:ComparisonSpectrum}
    \end{figure}

\begin{figure}[!htbp]
      \centering
      \includegraphics[width=0.7\linewidth]{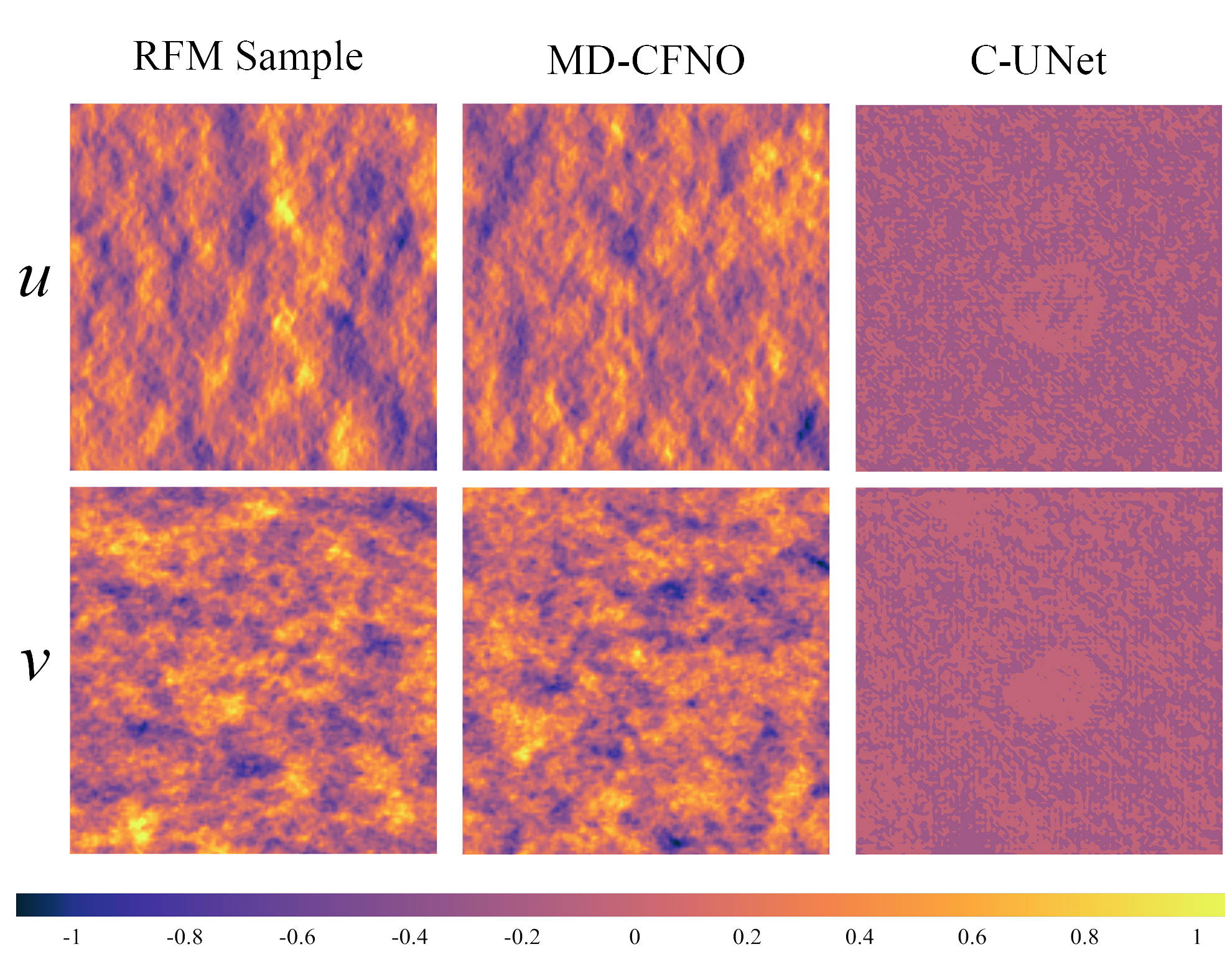}
      \caption{Visual comparison of instantaneous velocity field contours at the reference condition (Case 0).}
      \label{fig:ComparisonVelocity}
    \end{figure}

    A reference condition is selected at the center of the validation parameter space (Case~0: $\mathit{TKE}=0.1606~\mathrm{m^2/s^2}$, $\varepsilon=1.3032~\mathrm{m^2/s^3}$), and the proposed MD-CFNO is compared with a parameter-matched conditional U-Net (C-UNet). C-UNet is adopted as the baseline because its encoder-decoder structure with multi-scale skip connections has shown robust performance in field reconstruction and image-to-image generation \cite{isola2017image}, and it remains one of the most widely used convolutional architectures in physics-informed learning. The comparison results are shown in Fig.~\ref{fig:ComparisonSpectrum} and Fig.~\ref{fig:ComparisonVelocity}. The C-UNet baseline exhibits clear limitations. Primarily due to its constrained local receptive field and the lack of explicit wavenumber-domain constraints, it tends to regress to the mean. Since vortex locations vary randomly across realizations, the network output is biased toward the conditional mean of the posterior distribution, leading to overly smoothed fields. In the wavenumber domain (Fig.~\ref{fig:ComparisonSpectrum}), this behavior manifests as severe energy attenuation in the high-energy range, indicating loss of essential spectral content. In contrast, the proposed MD-CFNO leverages the global integral operator of Fourier layers in the wavenumber domain, alleviates these deficiencies, and recovers the energy distribution across scales, yielding a much closer agreement with the target spectrum than the C-UNet.
    
\begin{figure}[!htbp]
      \centering
      \includegraphics[width=0.5\linewidth]{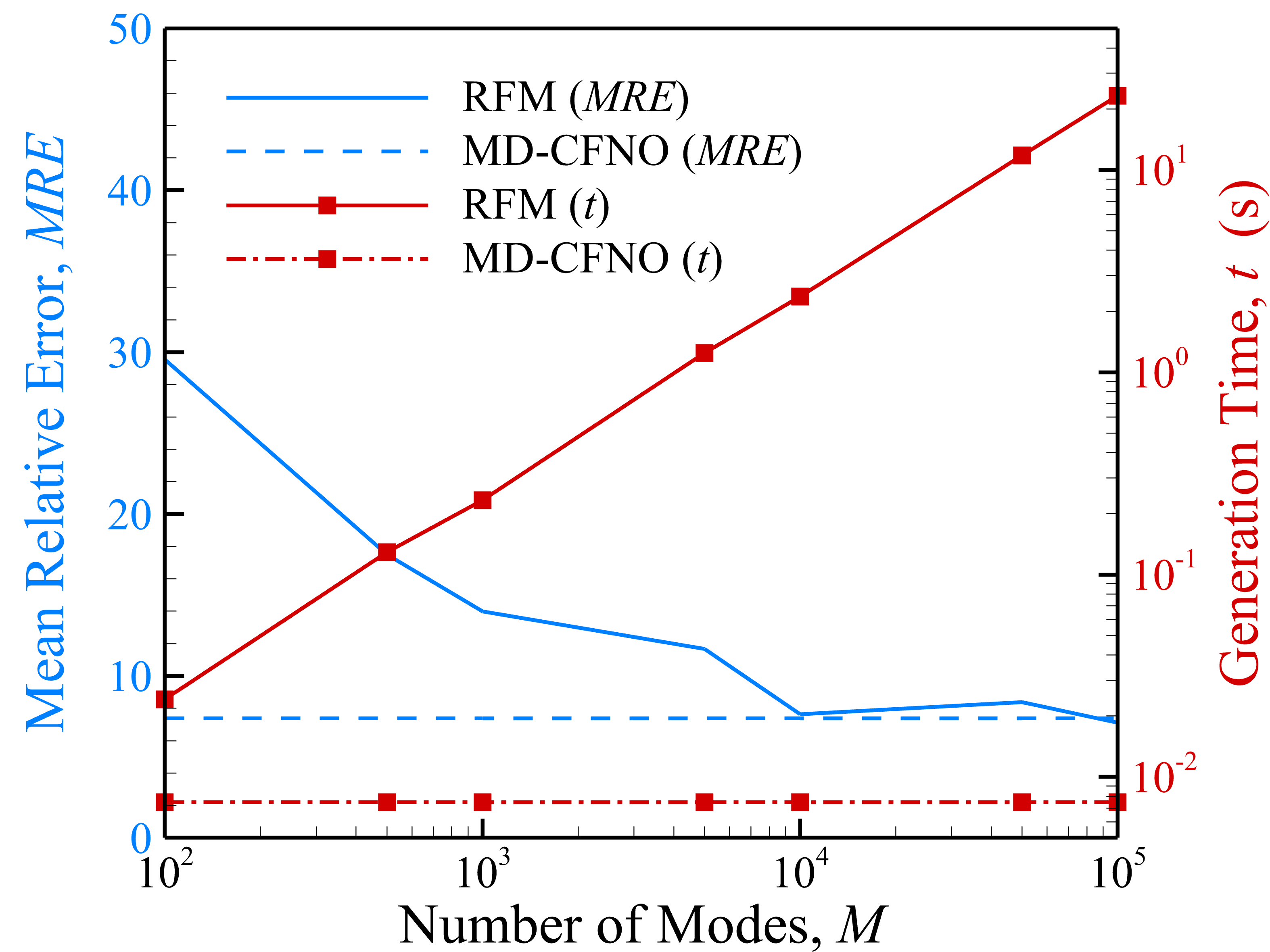}
      \caption{Quantitative comparison between the proposed MD-CFNO and RFM with varying mode numbers about computational efficiency and statistical accuracy.}
      \label{fig:time}
    \end{figure}

    The proposed MD-CFNO is further quantified against RFM, using RFM results as a reference. Following our previous study \cite{lin2025}, the mean relative error (MRE) of the energy spectrum over the full wavenumber range is used as the metric. As summarized in Table~\ref{tab:simple_comparison} and Fig.~\ref{fig:time}, the accuracy of RFM depends strongly on the number of modes $M$, while the computational cost grows approximately linearly with $M$. To reach a spectral accuracy comparable to the proposed MD-CFNO (MRE $\approx 7\%$), RFM requires $M \approx 10^4$, with an average single-sample generation time of about 2.36~s. In contrast, the proposed MD-CFNO decouples the inference cost from $M$ and attains a constant $O(1)$ inference complexity, generating one sample in 7.52~ms. At comparable spectral accuracy, this corresponds to an efficiency gain of approximately $300$.

    \begin{table}[!htbp]
        \caption{Quantitative performance comparison between the proposed MD-CFNO and the traditional RFM with varying mode numbers.}
        \label{tab:simple_comparison}
        \centering
        \small 

        \newcolumntype{Y}{>{\centering\arraybackslash}X}
    
        \begin{tabularx}{\textwidth}{c c Y Y Y Y}
            \toprule 
            Method & Modes & MRE (\%) & Generation Time (s) & Throughput (samples/s) & Time Ratio (RFM/MD-CFNO) \\
            \midrule 
            
            \textbf{MD-CFNO} & -- & 7.391 & 0.007515 & 133.1 & 1.0 \\
            
            \midrule 
            
            \multirow{7}{*}{\textbf{RFM}} 
              & $100$     & 29.56 & 0.02422 & 41.29   & 3.222 \\
              & $500$     & 17.49 & 0.12876 & 7.837   & 16.98 \\
              & $1000$    & 13.99 & 0.2334  & 4.285   & 31.06 \\
              & $5000$    & 11.66 & 1.245   & 0.8033  & 165.7 \\
              & $10000$   & 7.633 & 2.360   & 0.4238  & 314.0 \\
              & $50000$   & 8.368 & 11.76   & 0.08504 & 1565 \\
              & $100000$  & 7.117 & 23.24   & 0.04303  & 3092 \\
            \bottomrule 
        \end{tabularx}
    \end{table}

    %=================================================================================================================
    \subsection{Seed Generalization under Reference Condition}

\begin{figure}[!htbp]
      \centering
      \includegraphics[width=0.5\linewidth]{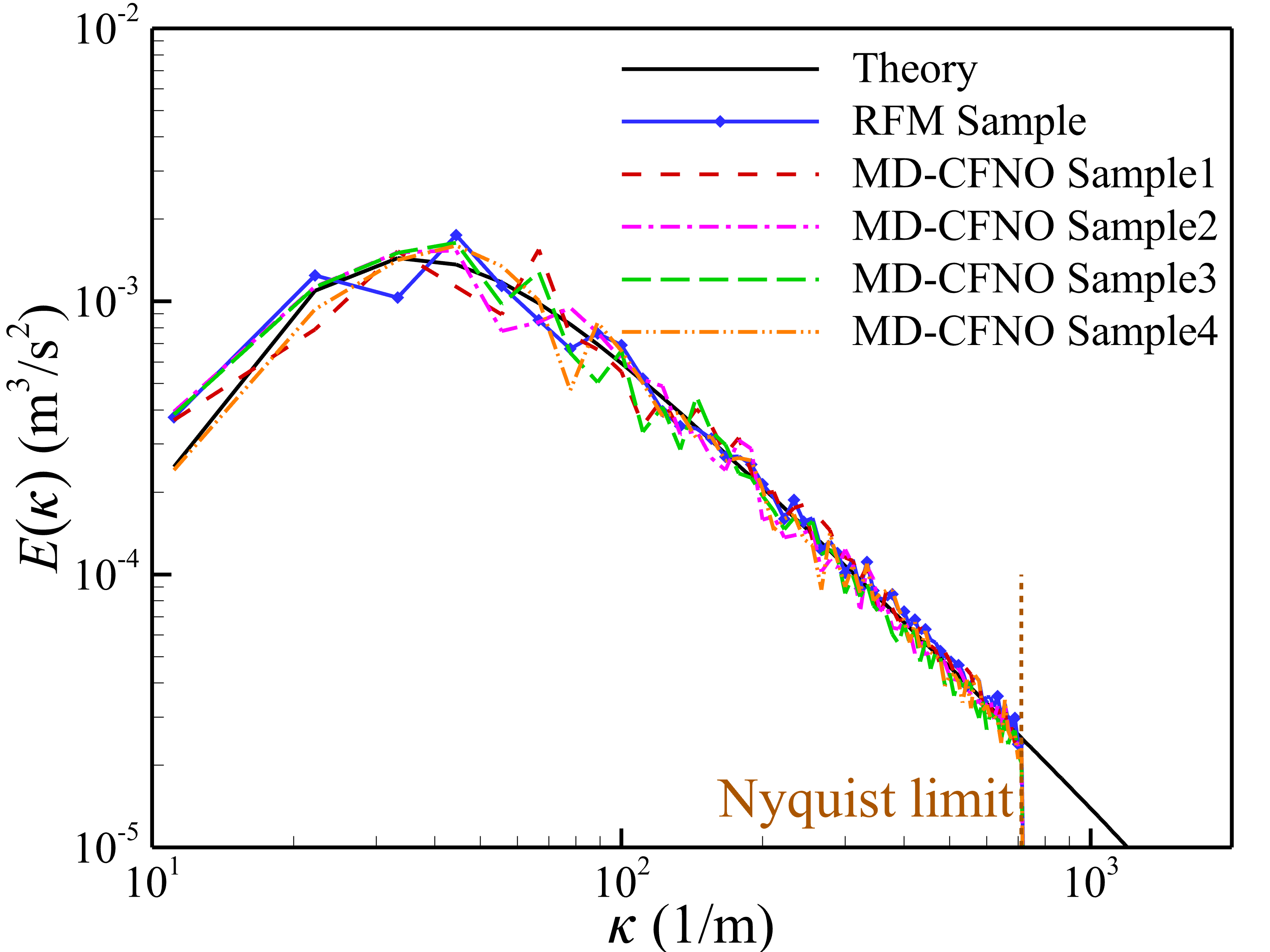}
      \caption{Verification of statistical consistency across varying random seeds for the reference condition (Case 0).}
      \label{fig:SameSpectrum}
    \end{figure}

\begin{figure}[!htbp]
      \centering
      \includegraphics[width=\linewidth]{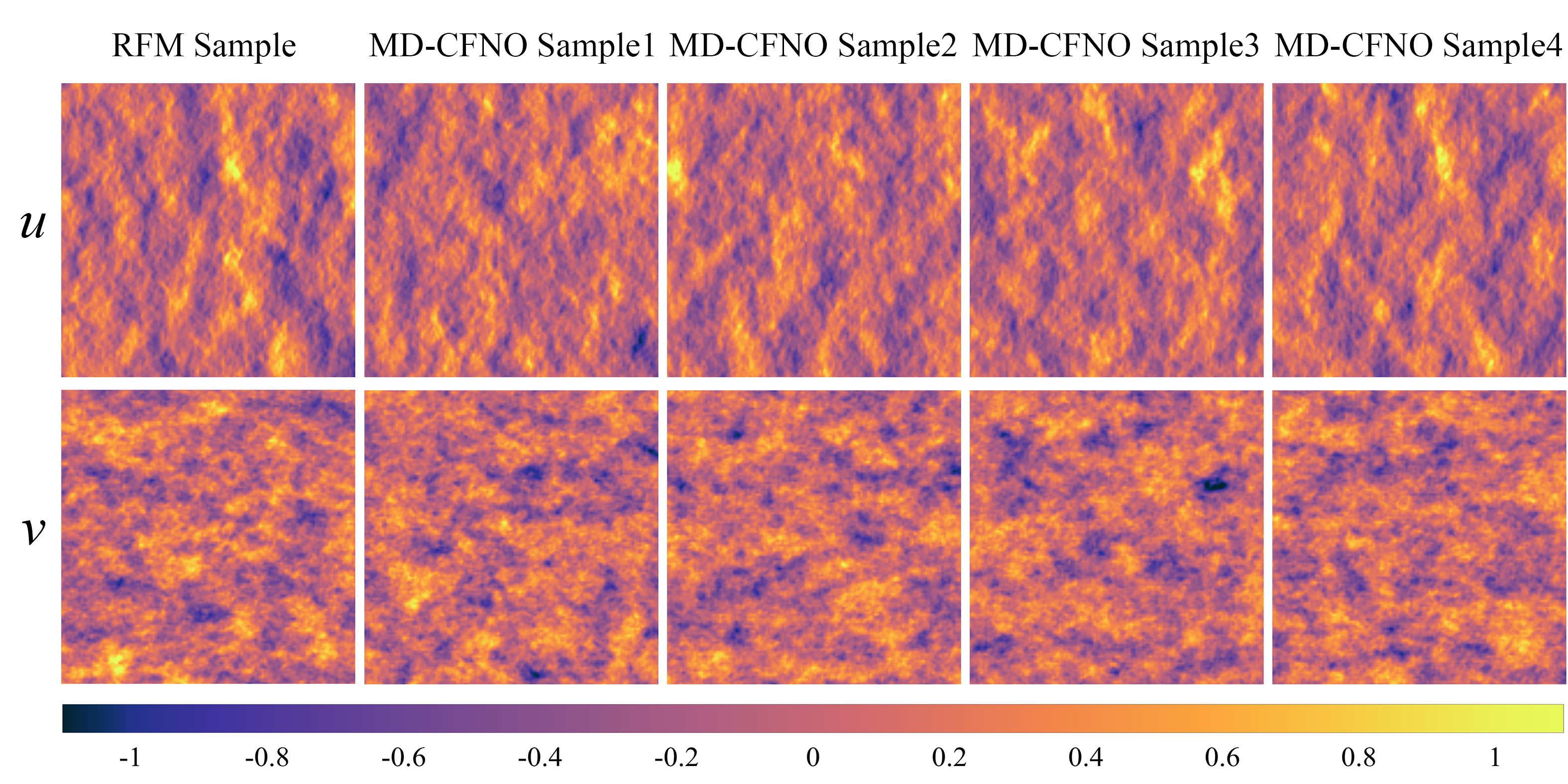}
      \caption{Diversity of generated flow structures under identical physical conditions (Case 0) driven by different random noise inputs.}
      \label{fig:Velocity}
    \end{figure}

    This section examines the stochastic generation capability and statistical consistency of the proposed MD-CFNO under a fixed physical condition (Case~0). The condition parameters $(\mathit{TKE}, \varepsilon)$ are held constant while only the input noise $z$ is varied to produce multiple realizations. Specifically, the proposed MD-CFNO is used to generate four fluctuating velocity-field samples, and the corresponding energy spectra are evaluated.
    As shown in Fig.~\ref{fig:SameSpectrum}, the spectra generated under different random seeds remain in close agreement with the prescribed theoretical spectrum, indicating that the model captures the underlying physical constraints rather than reproducing individual realizations. Instantaneous field visualizations in Fig.~\ref{fig:Velocity} further indicate that, while maintaining consistent macroscopic energy levels, the generated flow fields exhibit pronounced stochastic diversity in vortex locations and morphology.

    %=================================================================================================================
    \subsection{Parametric Generalization to Boundary Conditions}
    To rigorously assess generalization, four corner conditions of the parameter space (Cases~1--4; see Fig.~\ref{fig:SpectrumSampling}) are selected as an out-of-distribution extrapolation test set. These conditions lie on the boundary of the training distribution in the $(\mathit{TKE},\varepsilon)$ space and constitute an extrapolation regime, representing extreme combinations of $\mathit{TKE}$ and $\varepsilon$ that are excluded from training. Specifically, Case~1 corresponds to low $\mathit{TKE}$ and low $\varepsilon$ ($\mathit{TKE}=0.03~\mathrm{m}^2/\mathrm{s}^2$, $\varepsilon=0.20~\mathrm{m}^2/\mathrm{s}^3$); Case~2 corresponds to low $\mathit{TKE}$ and high $\varepsilon$ ($\mathit{TKE}=0.03~\mathrm{m}^2/\mathrm{s}^2$, $\varepsilon=2.00~\mathrm{m}^2/\mathrm{s}^3$); Case~3 corresponds to high $\mathit{TKE}$ and low $\varepsilon$ ($\mathit{TKE}=0.30~\mathrm{m}^2/\mathrm{s}^2$, $\varepsilon=0.20~\mathrm{m}^2/\mathrm{s}^3$); and Case~4 corresponds to high $\mathit{TKE}$ and high $\varepsilon$ ($\mathit{TKE}=0.30~\mathrm{m}^2/\mathrm{s}^2$, $\varepsilon=2.00~\mathrm{m}^2/\mathrm{s}^3$). These cases span distinct spectral shapes and turbulence energy levels, providing a stringent evaluation of robustness under out-of-distribution physical conditions.

\begin{figure}[!htb]
    \centering
    \begin{subfigure}[b]{0.48\linewidth}
        \centering
        \includegraphics[width=\linewidth]{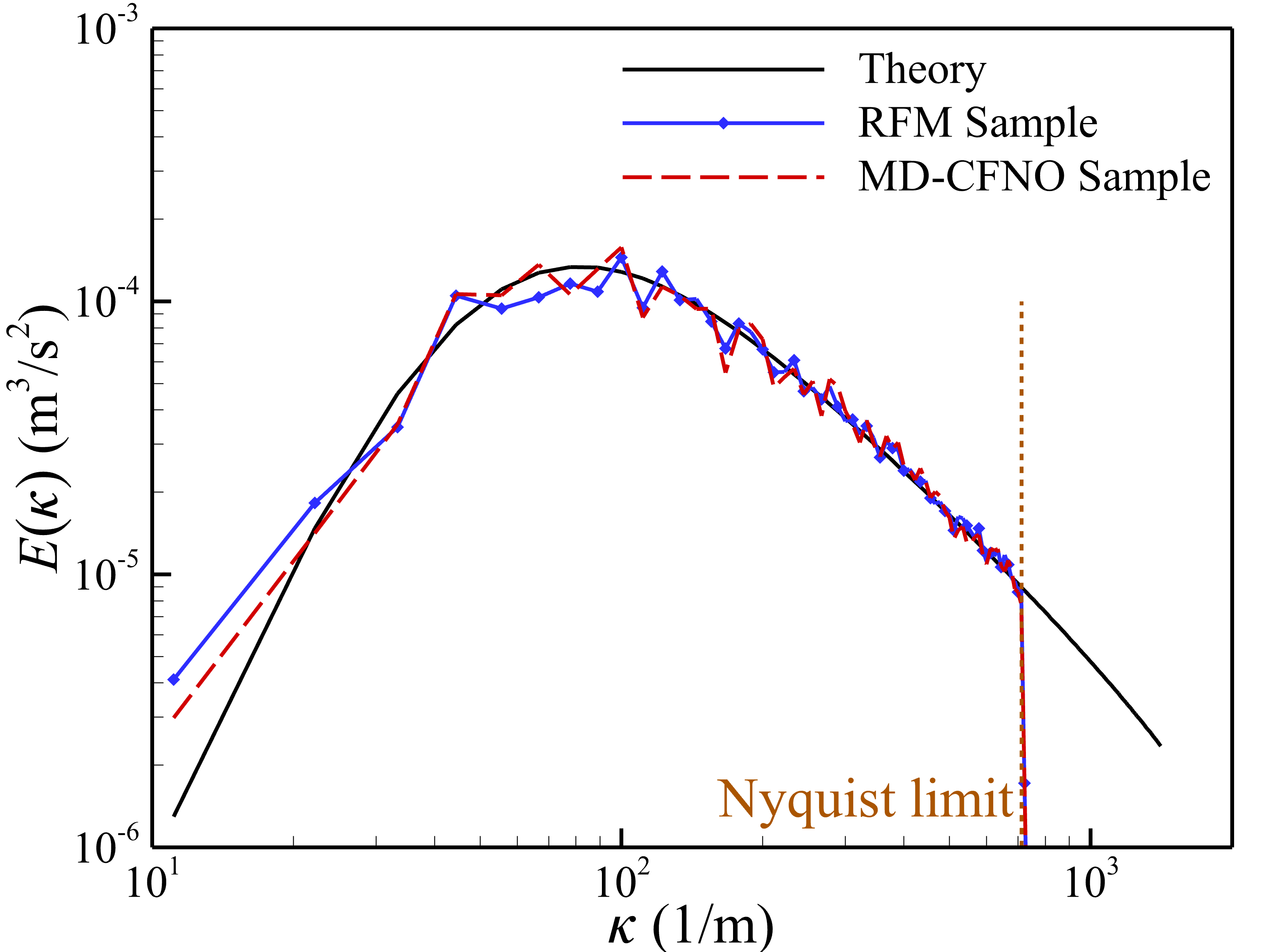}
        \caption{Case 1: $\mathit{TKE} = 0.03 \, \mathrm{m}^2/\mathrm{s}^2, \varepsilon = 0.20 \, \mathrm{m}^2/\mathrm{s}^3$.}
        \label{fig:cond1}
    \end{subfigure}
    \hfill
    \begin{subfigure}[b]{0.48\linewidth}
        \centering
        \includegraphics[width=\linewidth]{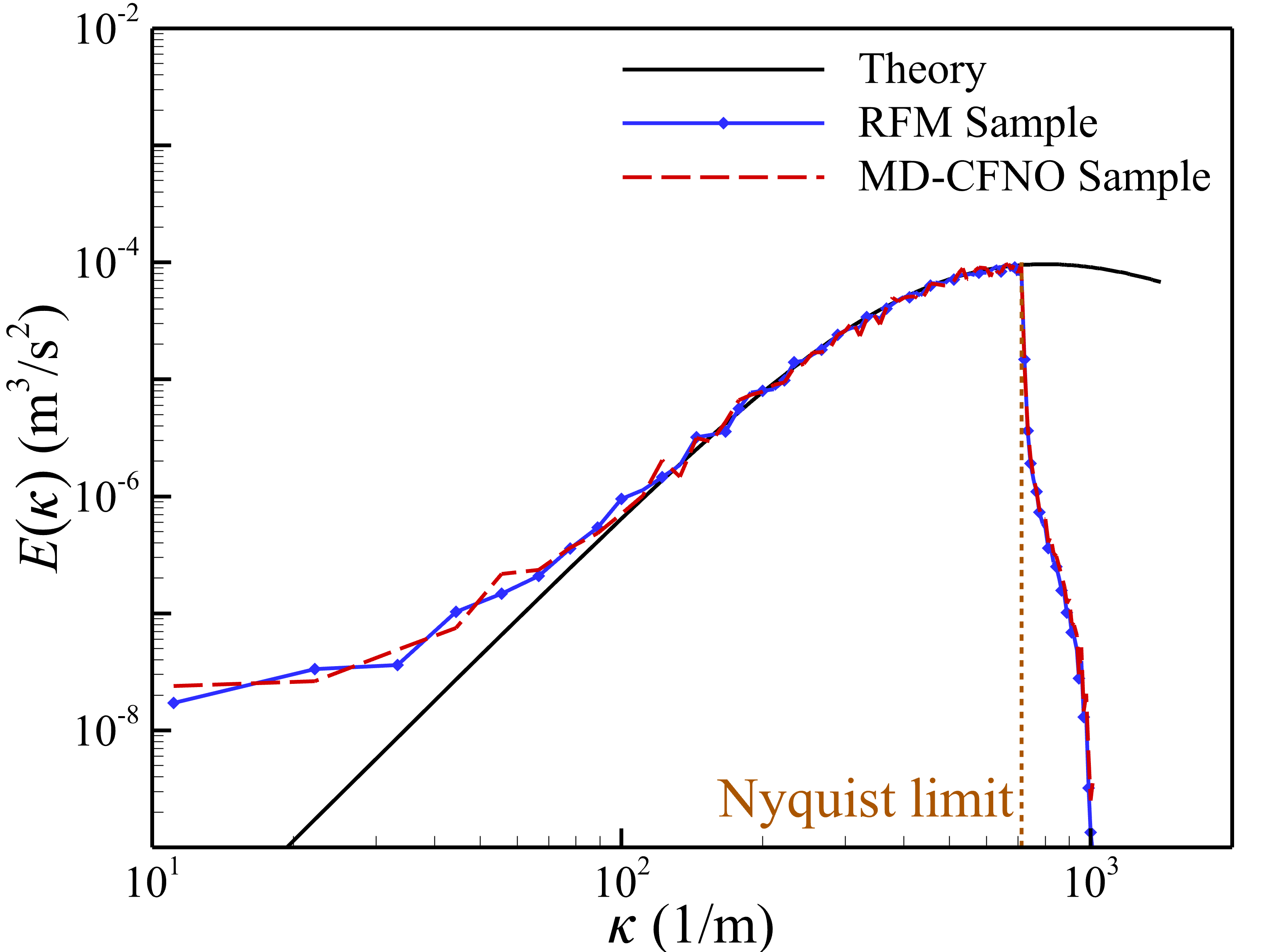}
        \caption{Case 2: $\mathit{TKE} = 0.03 \, \mathrm{m}^2/\mathrm{s}^2, \varepsilon = 2.00 \, \mathrm{m}^2/\mathrm{s}^3$.}
        \label{fig:cond2}
    \end{subfigure}
    \vspace{1em} 
    \begin{subfigure}[b]{0.48\linewidth}
        \centering
        \includegraphics[width=\linewidth]{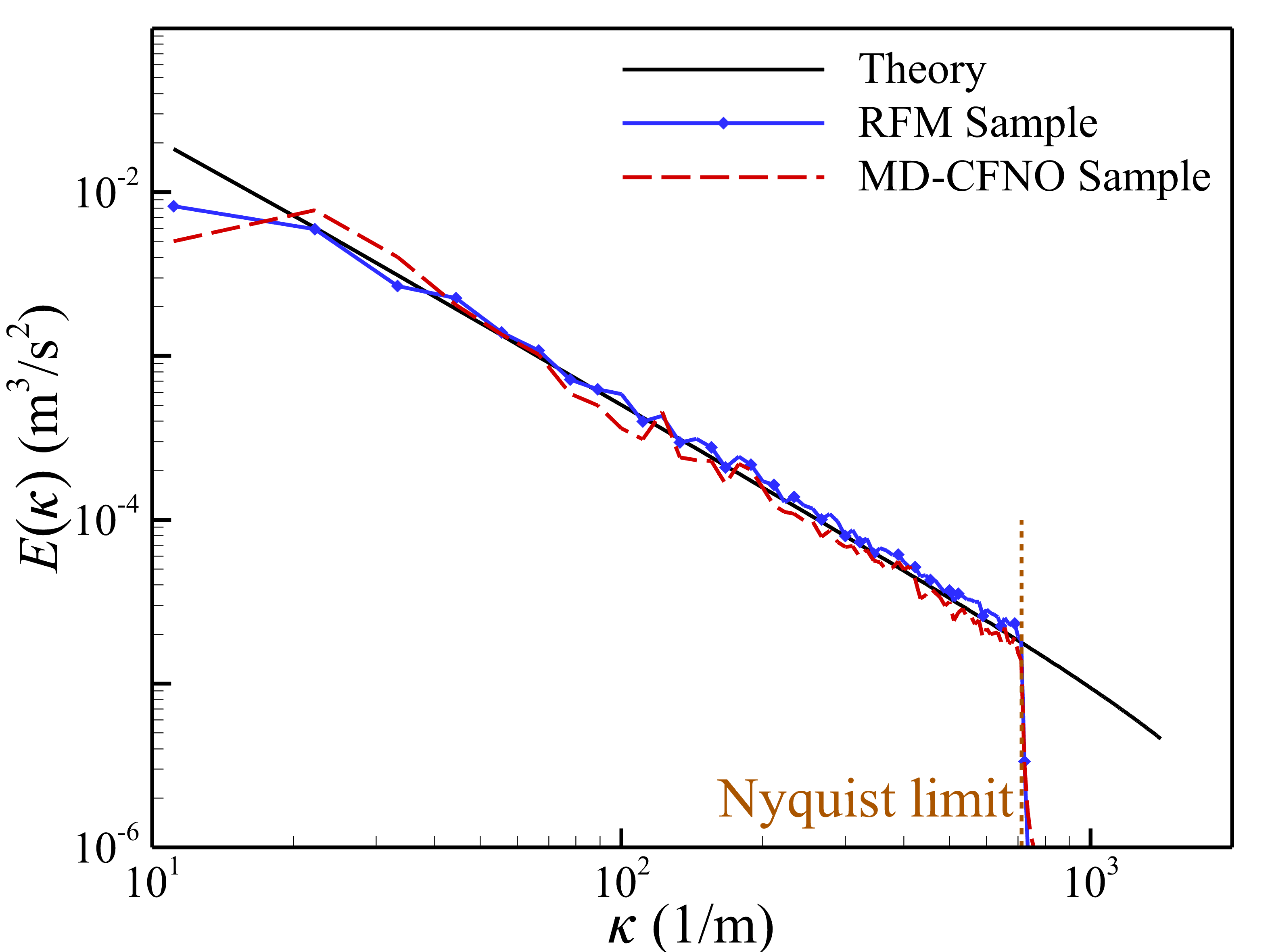}
        \caption{Case 3: $\mathit{TKE} = 0.30 \, \mathrm{m}^2/\mathrm{s}^2, \varepsilon = 0.20 \, \mathrm{m}^2/\mathrm{s}^3$.}
        \label{fig:cond3}
    \end{subfigure}
    \hfill
    \begin{subfigure}[b]{0.48\linewidth}
        \centering
        \includegraphics[width=\linewidth]{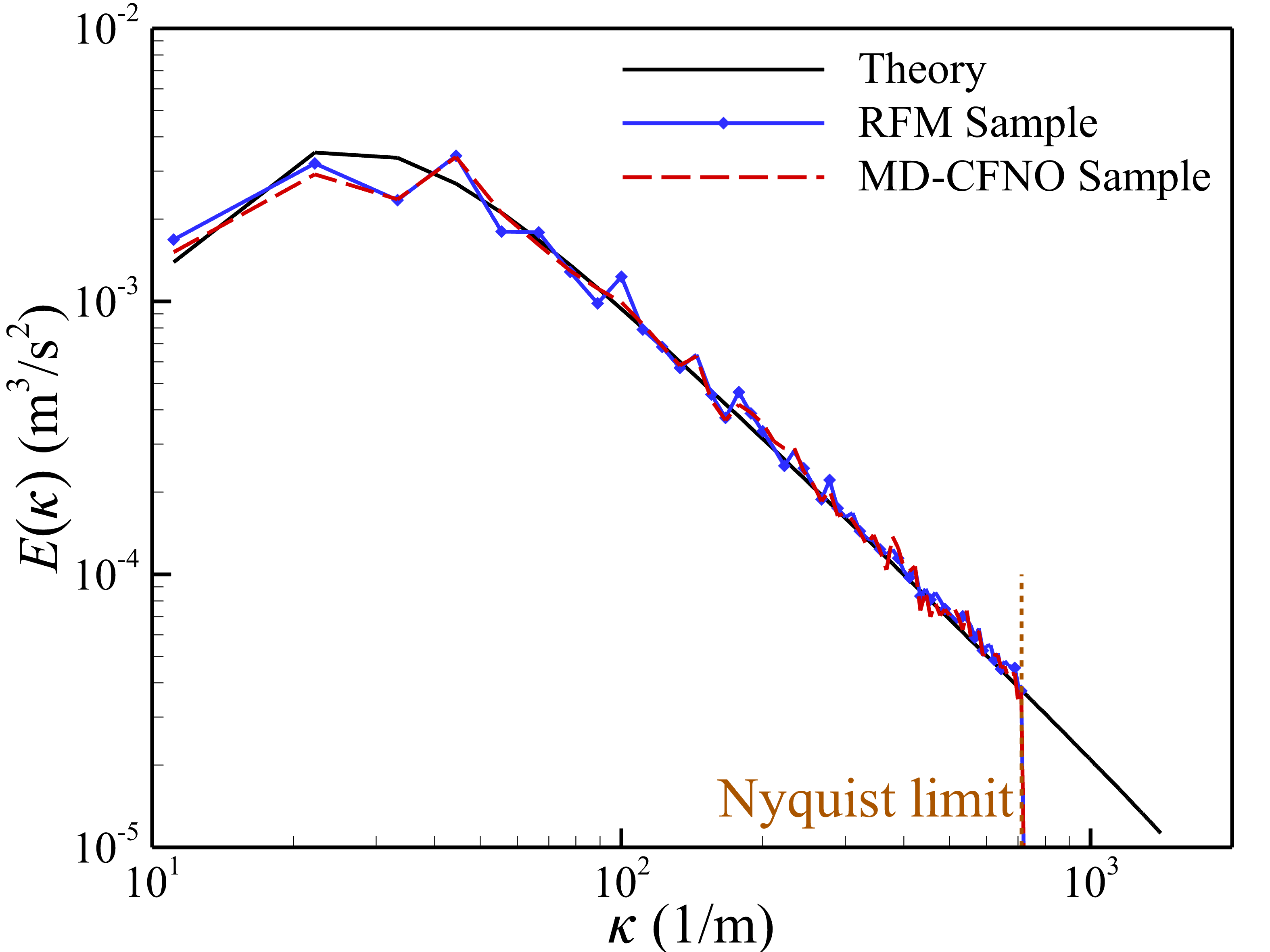}
        \caption{Case 4: $\mathit{TKE} = 0.30 \, \mathrm{m}^2/\mathrm{s}^2, \varepsilon = 2.00 \, \mathrm{m}^2/\mathrm{s}^3$.}
        \label{fig:cond4}
    \end{subfigure}
    \caption{Assessment of extrapolation capabilities on the selected generalization cases.}
    \label{fig:CondComp}
    \end{figure}

    Figure~\ref{fig:CondComp} shows that, for all test conditions, the spectra predicted by the proposed MD-CFNO closely match the theoretical spectra. Comparisons across cases indicate that variations in $\mathit{TKE}$ primarily induce a vertical shift of the spectrum magnitude, whereas variations in $\varepsilon$ mainly affect the high-wavenumber decay and the cutoff location. The proposed MD-CFNO captures these condition-induced spectral changes and does not exhibit nonphysical energy pile-up near the Nyquist limit. These results indicate that the model learns the mapping between conditioning parameters and spectral distributions, rather than producing a generic or averaged spectrum, and adapts the generated fields according to the input physical parameters. Since these test conditions, together with the entire parameter-space boundary, are excluded from training, the results further indicate strong out-of-distribution generalization in the extrapolation regime at the edge of the parameter space.
    
    %=================================================================================================================
    \subsection{Impact of Loss Function}
    \label{subsec:Ablation}

\begin{figure}[!htbp]
      \centering
      \includegraphics[width=0.5\linewidth]{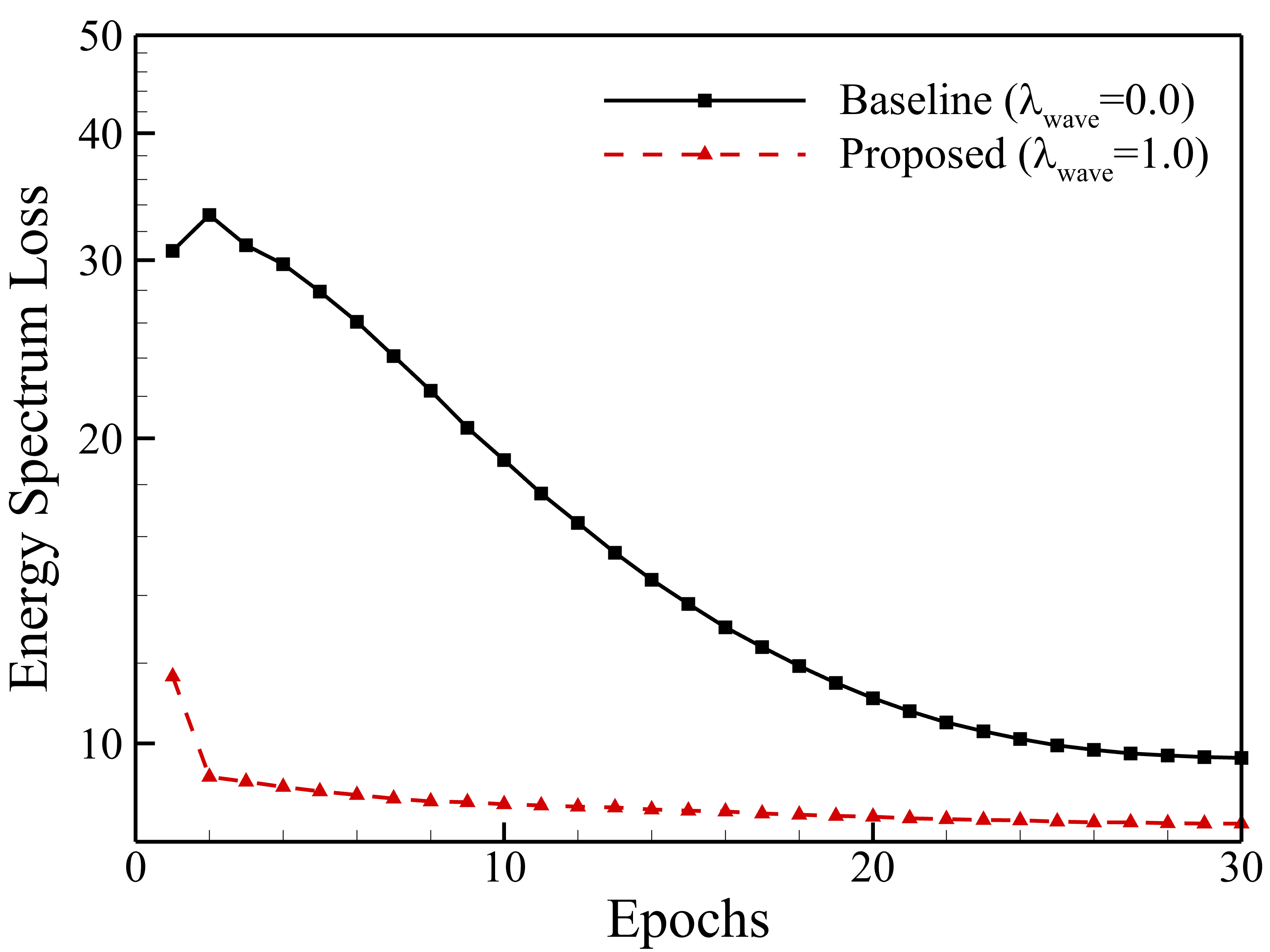}
      \caption{Impact of the wavenumber-domain loss term on training convergence.}
      \label{fig:LossComparison}
    \end{figure}

     This section investigates the role of the wavenumber-domain constraint term $\mathcal{L}_{\mathrm{wave}}$ in improving training convergence. With the network architecture and all other hyperparameters kept identical, two settings are compared: a baseline that uses only the spatial-domain reconstruction loss (Baseline, $\lambda_{\mathrm{wave}}=0.0$) and a configuration that additionally includes the wavenumber-domain constraint (Proposed, $\lambda_{\mathrm{wave}}=1.0$). Figure~\ref{fig:LossComparison} reports the evolution of the spectrum consistency loss in Eq.~\ref{eq:loss_spec} over training epochs for both settings. A clear difference is observed. With $\mathcal{L}_{\mathrm{wave}}$ included, the spectral loss decreases rapidly at the early stage of training and reaches a low level within a few epochs, whereas the baseline shows a slower and more delayed decrease. This indicates that the wavenumber-domain constraint provides a more informative gradient signal that accelerates optimization toward the target spectral distribution. At the end of training (30 epochs), the model trained with $\mathcal{L}_{\mathrm{wave}}$ also achieves a lower spectral reconstruction loss, supporting the necessity of explicit wavenumber-domain supervision for spectrum-consistent synthetic turbulence generation.     

%=================================================================================================================
\section{Conclusions}
\label{sec:5-Conclusion}
   
    This paper proposes a model-driven conditional Fourier neural operator (MD-CFNO) for generating spectrum-consistent synthetic turbulent velocity fields.
    To mitigate the regression-to-the-mean behavior commonly observed in spatial-domain networks, the conditional Fourier neural operator leverages global wavenumber-domain integration to better preserve spectrum consistency.
    Using a model-driven strategy, the proposed MD-CFNO reduces reliance on expensive direct numerical simulation data and supports training over an integral-scale Reynolds-number regime spanning 45-45,000, with scalability to higher regimes limited only by grid resolution.
    Furthermore, incorporating explicit wavenumber-domain constraints accelerates convergence and promotes closer agreement with theoretical energy spectra.
    Systematic evaluations under interpolation regimes and boundary extrapolation settings show that the proposed MD-CFNO enables controllable synthesis while maintaining spectrum consistency. 
    
    While the present study relies on a baseline random Fourier model that mainly represents homogeneous isotropic turbulence, the proposed MD-CFNO is not limited to this setting. Future work will incorporate more expressive random Fourier model variants to extend the proposed MD-CFNO to complex anisotropic flows and spatiotemporal evolution.   

% Appendixes ------------------------------------------------------------------
\section*{Acknowledgments}
  This work is supported by NSFC the Excellence Research Group Program for multiscale problems in nonlinear mechanics (Grant No. 12588201), the Chinese Academy of Sciences Project for Young Scientists in Basic Research (Grant No. YSBR-087), and the Strategic Priority Research Program of Chinese Academy of Sciences (Grant No. XDB0620102).

\section*{Data and Code Availability}
  The source code for the MD-CFNO model are openly available on GitHub at \url{https://github.com/Hongyuan0/CFNO_Turbulence}. The datasets generated and analyzed during the current study are available from the corresponding author on reasonable request.

\section*{Declaration of generative AI and AI-assisted technologies in the manuscript preparation process}
  During the preparation of this work the authors used Gemini in order to improve readability, grammar, and language. After using this tool, the authors reviewed and edited the content as needed and take full responsibility for the content of the published article.

% References ----------------------------------------------------------------
% \FloatBarrier
%\bibliographystyle{elsarticle-harv}
% \bibliographystyle{elsarticle-num-names}
\bibliographystyle{elsarticle-num}
\bibliography{refs}

\end{CJK*}  % chinese
\end{document}